\documentclass[12pt,groupedaddress,superscriptaddress,showkeys,showpacs,citeautoscript]{revtex4}
\usepackage{amssymb}
\usepackage{graphicx}

\begin{document}
\title{Origin of Periodic Modulations in the Transient Reflectivity Signal at Cryogenic Temperatures}

\author{Salahuddin Khan}
\email{skhan@rrcat.gov.in}
\author{Rama Chari} \author{J. Jayabalan}
\affiliation{Laser Physics Applications Section, Raja Ramanna Centre for
Advanced Technology, Indore, India.}
\author{Suparna Pal}\author{T. K. Sharma}
\affiliation{Solid State Laser Division, Raja Ramanna Centre for Advanced
Technology, Indore, India.}
\author{A. K. Sagar}\author{M. S. Ansari}\author{P. K. Kush}
\affiliation{Cryo-engineering and Cryo-module Development Section, Raja Ramanna
Centre for Advanced Technology, Indore, India.}

\begin{abstract}
Periodic modulations that appear in the low-temperature transient reflectivity
signal of a GaAsP/AlGaAs single quantum well is studied. Similar anomalous
oscillations are also observed in layered manganite [K.~Kouyama  et.al. J.
Phys. Soc. Jpn. 76:123702(1--3), 2007]. We show that such periodic modulations
are caused by changes in the linear reflectivity of the sample
during transient reflectivity measurements. Studied carried out on reflectivity of
different materials under identical conditions shows that these modulations on
the true transient reflectivity signal are caused by condensation
of residual gases on the surface of quantum well. Methods to obtain reliable
transient reflectivity data are also described.
\end{abstract}

\date{\today}

\maketitle

\section{Introduction}
Femtosecond pump-probe spectroscopy is used for studying the sub-picosecond carrier
response of materials which are to be used in devices like lasers, modulators, detectors
etc \cite{rsiexp, ultrafast, laser}. For nanostructured materials grown on opaque substrate
in optoelectronic devices, very often it becomes necessary to do the pump-probe measurements in a
reflective geometry \cite{nanolettqdtr,TR2,TR}. In such measurements the transient
change in reflectivity, $\Delta R/R$ can be as low as $\sim$ 10$^{-6}$.
Similar to transient reflectivity several other well known techniques
like photo-modulated reflectivity and ellipsometry also use reflection geometry \cite{pr,insituellipsometry}.
Hence to get reliable data from these
techniques it is essential to maintain the quality and purity of the sample surface. In order to
achieve this good surface quality cleaving or etching of the surface is done just
before the experiment or in-situ measurements are carried out.

In the pump-probe technique, transient reflectivity measurement of a sample can take
few tens of minutes for a complete a scan. Thus, even a slow surface contamination
process can modulate the actual signal originating from the ultrafast response of the sample.
Transient reflectivity measurements carried out at low temperatures on
$La_{0.5}Sr_{1.5}MnO$ showed oscillations in the signal with an anomalously
long time period which is not well understood \cite{oscillationsmanganite}.
Similar periodic modulations were also reported in reflectivity measurements on bulk
ZnO \cite{Znosurfacecharge}. Our transient reflectivity measurements on GaAsP/AlGaAs
quantum wells at low temperatures showed similar modulations in the transient reflectivity
signal. By studying reflectivity of different materials under similar
experimental conditions as that of quantum wells we show that these
modulations on the actual transient reflectivity signal are caused by condensation
of residual gases on the surface of quantum well. We also show the various precautionary
measures which can be taken to obtain reliable transient reflectivity data.

\section{Experiment}
The transient reflectivity measurements were carried out in a standard degenerate
pump-probe geometry with a 90 fs, 82 MHz Ti:Saphire laser \cite{TR2,TR}.
The laser was operated at peak wavelength of 786 nm during transient reflectivity
measurements. The samples under study were mounted in a closed-cycle cryostat, which was
evacuated by a diffusion pump connected through a liquid nitrogen trap \cite{cryo-cooler}.
The sample temperature was kept stable within 0.1 K during the experiments. The residual
pressure inside the  cryostat was of the order of $\sim$ $10^{-6}$ millibar. The pump
and probe beams were made to fall on the sample kept inside the cryostat through
an one inch optical window. The angle of incidence of pump and probe beams
on the sample were $\sim$ 0$^0$ and $10^{0}$ respectively.
The pump beam was mechanically chopped and the reflected power of the probe beam
from the sample was detected by a photodiode and lock-in-amplifier combination.
The change in reflectivity of probe pulse caused by the pump
pulse was then recorded at different pump-probe delays. The complete evolution of the
transient reflectivity during the first few hundreds of picosecond after excitation
by the pump pulse was built up by taking such repeated observations at different
pump-probe delays. This process of data accumulation takes a few tens of minutes
to complete. The samples used in our transient reflectivity studies was a GaAsP/AlGaAs
single quantum well (QW). Quantum well structure was formed by growing a
GaAs$_{0.86}$P$_{0.14}$/Al$_{0.7}$Ga$_{0.3}$As in between Al$_{0.7}$Ga$_{0.3}$As
barriers of thicknesses 250 nm and 50 nm thickness on the bottom and top sides. The
QW structure is deposited on a [001] n+ doped GaAs substrate \cite{qw}.

\section{Results and Discussion}

\begin{figure}
\includegraphics[width=0.85\textwidth]{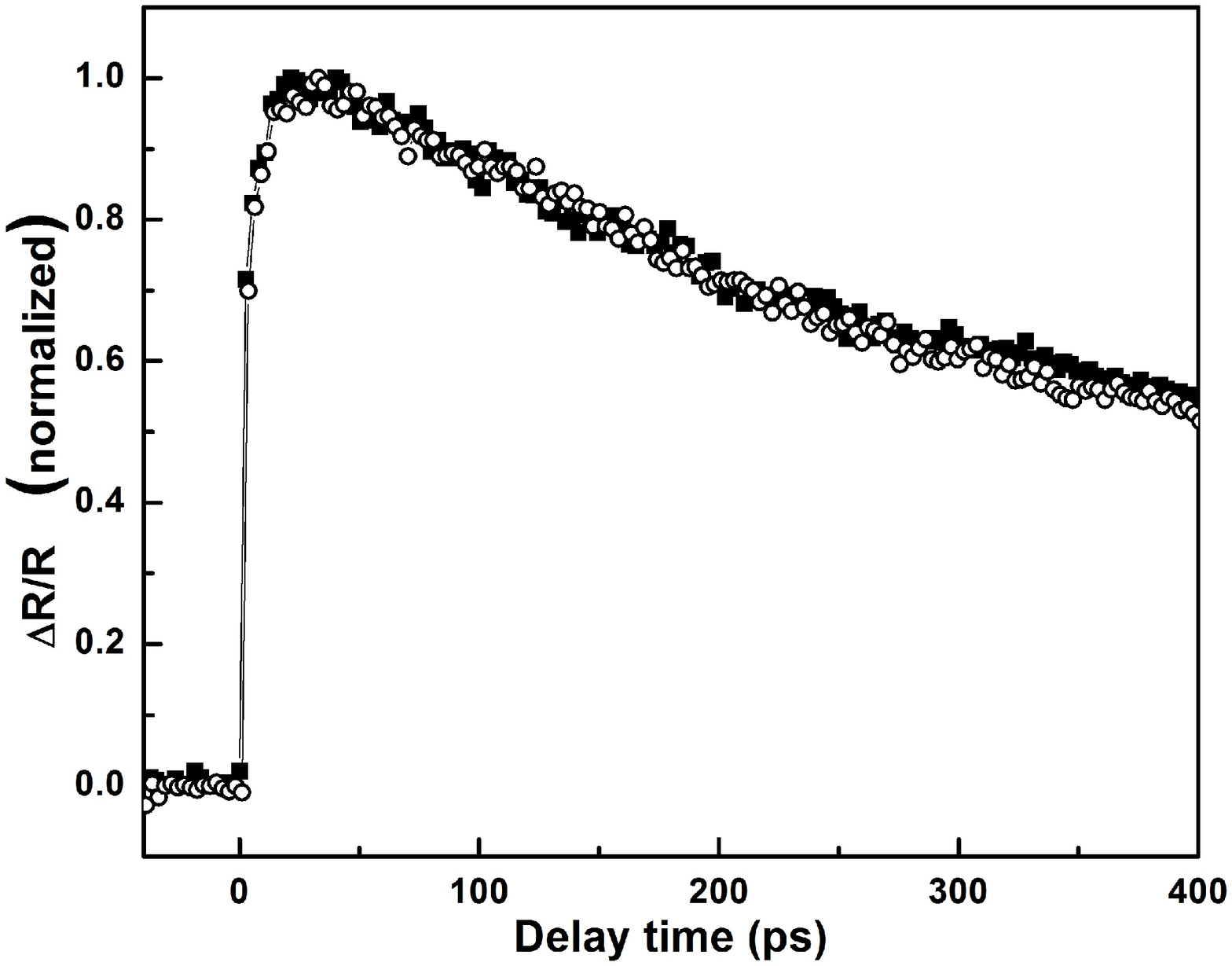}
\caption{\label{fig:Fig1} Two representative scans of transient reflectivity signal
from GaAsP/AlGaAs quantum well at room temperature (300 K) under identical conditions.
The data was always repeatable within the experimental errors.}
\end{figure}

\begin{figure}
\includegraphics[width=0.85\textwidth]{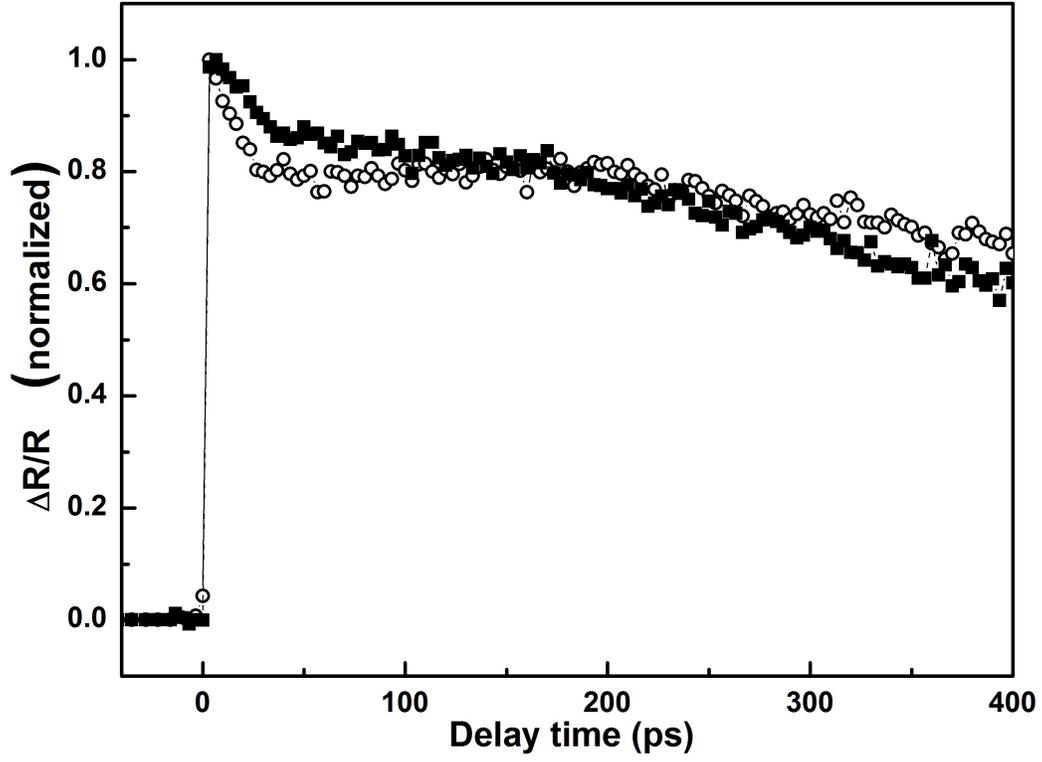}
\caption{\label{fig:Fig2} Two representative scans of transient reflectivity signal
($\Delta R / R$) of GaAsP/AlGaAs quantum well at 50 K. The time evolution of the
signal was different on each scans.}
\end{figure}

Figure \ref{fig:Fig1} shows the transient reflectivity signal from GaAsP/AlGaAs quantum well
measured at room temperature (300 K). The 786 nm wavelength of the laser excites
photocarriers in the quantum well as well as the GaAs substrate. These photo-excited carriers
modifies the refractive index, and hence the reflectivity of the quantum well structure. This
results in the fast initial rise with time constant nearly 6 ps.
This initial fast change in the reflectivity recovers in $\sim$ 630 ps due to the
relaxation and transport of the carriers. Repeated measurements
under identical sample conditions shows that the transient reflectivity signal
remains same within the experimental errors (Fig. \ref{fig:Fig1}). However, the low temperature
transient reflectivity shows a very different behavior. Figure \ref{fig:Fig2} shows
the transient reflectivity of the same sample measured at the same experimental conditions
but at 50 K. Finer analysis of the data shows that the low temperature transient reflectivity
signal was similar in nature as that of room temperature but with additional oscillatory
modulations on it. Transient reflectivity measurements on $La_{0.5}Sr_{1.5}MnO$ at low
temperatures showed oscillations in the signal similar to that reported here \cite{oscillationsmanganite}.
However, they have reported that the origin of these
oscillations in the transient reflectivity signal is not clear yet. Repeated
measurement of transient reflectivity of the QW sample shows that the phase of
the oscillations in the signal kept changing on each scan. This causes the transient
reflectivity profile to change markedly in each scan (Fig. \ref{fig:Fig2}). In fact
we find that when the temperature of the sample was below 200 K the transient reflectivity
signal becomes not repeatable. The linear reflectivity of the QW sample measured by
using the probe pulse and by blocking the pump beam, itself showed such periodic
oscillations. Due to such oscillations, if the linear reflectivity measurement
starts at an arbitrary time the initial phase of the oscillation is different.
These modulations in the linear reflectivity of the sample causes the transient
reflectivity signal to change at each scan.

\begin{figure}
\includegraphics[width=0.85\textwidth]{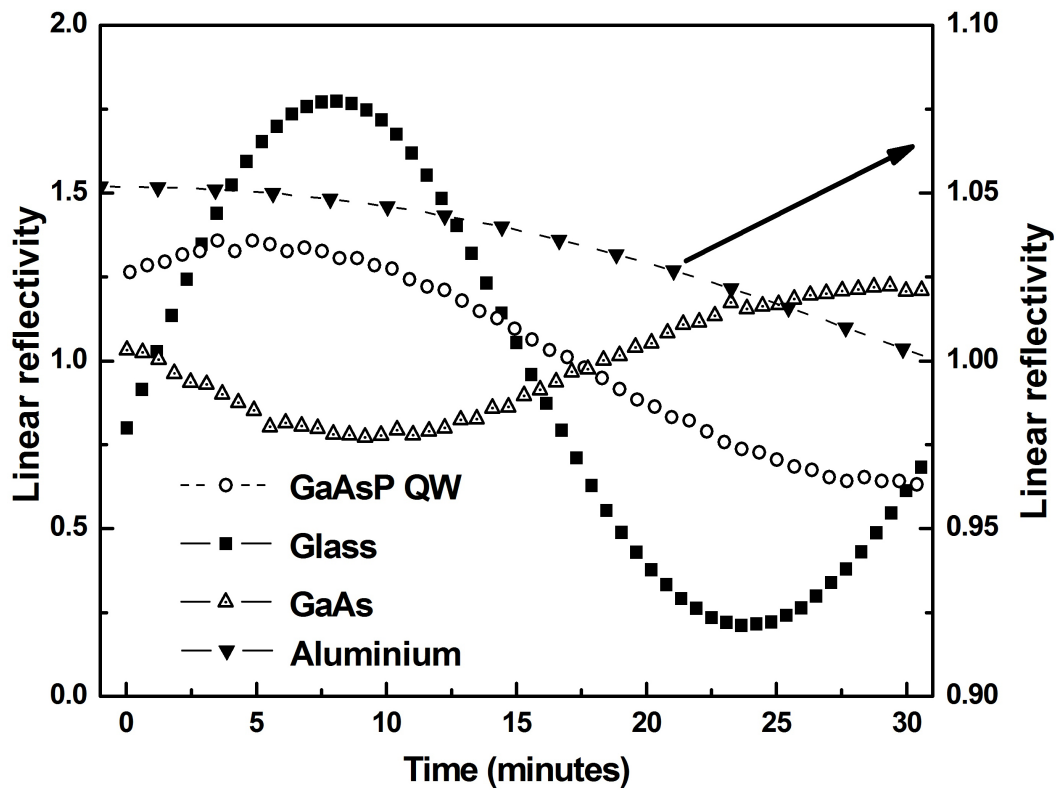}
\caption{\label{fig:Fig3} The variations in the linear reflectivity of various materials
at 50 K (wavelength of laser 632.8 nm). For ease of viewing, the reflected signal has
been normalized by dividing it with its corresponding mean value.}
\end{figure}

In order to understand the variations in the linear reflectivity of the
samples and to avoid high intensity effects we have performed the same
linear reflectivity measurement at 632.8 nm using a continuous wave low-power
He-Ne laser. Figure \ref{fig:Fig3} shows the variations measured in the linear
reflectivity of the QW sample at 50 K. The measured linear reflectivity of the
sample clearly shows an oscillatory behavior which was not noticeable at sample
temperatures above 200 K. These periodic modulations on the signal was similar in
nature to that reported for bulk ZnO \cite{Znosurfacecharge}. These oscillations in the
linear reflectivity in their case has been attributed to space change effect.
To identify the cause for the observed changes in the linear reflectivity our QW sample,
a material dependence study has been carried out on various samples, a glass plate
(insulator), a bulk GaAs wafer (semiconductor) and a bulk Aluminium block (metal).
The surface of these bulk materials were of optical quality.
The variations in the linear reflectivity of these samples are also shown in
Fig.\ref{fig:Fig3}. The oscillations in the linear reflectivity of GaAs wafer
remained nearly same in frequency and magnitude as that of the QW sample. The
magnitude of oscillation was found to be largest for glass. The observation of
similar periodic variations in the linear reflectivity of different kinds of materials
implies that the oscillations has a common origin independent of materials. We attribute the
observed oscillations in the reflectivity of the sample to condensation of water
vapor, residual gases and organic impurities present in the vacuum chamber on the
surface of the sample at low temperatures. As the sample cools, a thin film
of condensate starts growing on its surface. The resulting interference between the
reflection from the top and bottom interfaces of the growing film can give rise
to the observed modulations in the measured linear reflectivity oscillations.
This would also explain why similar oscillations are observed in all the materials under study.

In order to estimate the growth rate and the refractive index of the condensate material
we model the effect of a continuously growing thin film on the reflectivity of the sample.
Let $l$ be the thickness of the thin film growing on the sample at time $t$.
The net reflectivity of the film and the sample top surface at any given time is
\cite{Reflectcalculation,Reflectcalculation2}
\begin{equation}
R = \frac{{R_1  + \Omega^2 R_2  + 2 \Omega \sqrt {R_1 R_2 } \cos \left(4\pi n'l/\lambda \right)}}
{{1 + \Omega^2 R_1 R_2  + 2 \Omega \sqrt {R_1 R_2 } \cos \left( 4\pi n'l/\lambda \right)}},
\label{Eq:ThinFilm}
\end{equation}
where $R_1$ is the reflectivity of the air and thin film interface and $R_2$ is
the reflectivity of the thin film and sample interface. $n'$ and $n$ are the
refractive index of the thin film and sample materials at the wavelength of the
light used \cite{handbook,handbook2}. The absorption loss per pass $\Omega$ is $exp{(-\alpha l)}$, where $\alpha$
is the linear absorption coefficient of the thin film material.
Figure \ref{fig:Fig7} shows the calculated reflectivity oscillations
using Eq.\ref{Eq:ThinFilm} for 632.8 nm wavelength for different values
of $n'$, $g$ and $\alpha$. For a given growth rate and $n'$ of a non-absorbing thin
film, the period as well as the magnitude of oscillations remains constant with time.
This can be explained as follows: if either the sample or the film is
a poor reflector, then $R_1$$R_2$ $\ll$ 1 and the denominator can be taken to be
nearly 1 with this approximation and assuming that the film thickness increases
linearly with time the above equation written as
\begin{eqnarray}
R(t) =  A +  B \cos \left[\frac{4 \pi n' gt}{ \lambda} \right],
\end{eqnarray}
where $g = dl/dt$ is the growth rate of the film and $A = R_1 + R_2$ and $B =
2\sqrt{R_1 R_2}$. Thus, a non-absorbing thin film with constant growth rate
and $n'$ will have constant amplitude as well as constant time period.
Assuming that the growing thin film on our samples to
have constant growth rate, constant refractive index and no absorption,
we fit the experimental data using Eq.\ref{Eq:ThinFilm} to have an estimate for
the refractive index and growth rate of the film. Figure \ref{fig:Fig8} shows
linear reflectivity of measured from the GaAs along with its best fit using
Eq.\ref{Eq:ThinFilm}. Table \ref{t1} shows the estimated refractive index
and growth rate of the thin films condensing on different samples derived from
these fittings. In case of cryo-vacuum condensation of gases,
the crystalline structure, and density of the condensed film depends
on temperature, ambient atmosphere \cite{condensation,condensation1}. The
material and surface quality of the sample on which the gases are condensing
will also effect the growth rate and refractive index of the thin film.
Thus, the estimated refractive index and growth rates of the films are expected
to vary with the underlying sample.

Figure \ref{fig:Fig4} shows the measured time dependence of linear reflectivity
measured from the glass surface over a long observation period.
Although the variations in the period of oscillations
and the magnitude can be neglected over one cycle, the long time observation
clearly shows an increase in the time period as well as changes in the
magnitude of oscillations. Similar changes in the period of oscillations as well
as the magnitude has been found for all the samples. In Fig.\ref{fig:Fig7}, we
have also shown the simulated changes in the reflectivity of the sample-thin film
structure for the two cases: first the film has finite absorption and and second
the growth rate decreases linearly with time. The observed changes in the
magnitude and period of oscillations in the linear reflectivity of the
sample shown in Fig.\ref{fig:Fig4}, can be explained if we assume the thin film
to have finite absorption and reduction in growth rate as it grows.
It is expected that as the condensation starts, the amount of residual gas present
in the chamber reduces and this can lead to reduction in the growth rate of the
film. Thus, the observed oscillatory behavior in the linear
reflectivity is consistent with our model of condensation of a thin film on the
samples surface. In such thin film formation the growth rate should depend
strongly on the temperature of the sample. It is expected that as the temperature
of the sample increases the growth rate should reduce and hence the period of
oscillations should increase. Figure \ref{fig:Fig5} shows the temperature
dependence of period of oscillations on the reflectivity of GaAs sample.
Clearly the time period of reflectivity oscillations increases with
increase in temperature, supporting the formation of thin film on
the material under study.

Having understood the cause of modulations in the linear reflectivity, it is
essential to suppress these oscillations to get the proper transient reflectivity
signal. In order to reduce the amount of residual gases,
the cryostat was conditioned by purging with nitrogen
and baking under vacuum. The reflectivity under various stages of conditioning
is shown in Fig.\ref{fig:Fig9}. Purging with nitrogen removes some of the residues
this results in increase of the time period of oscillation in the reflectivity.
However with a vacuum level of $1\times10^{-6}$ millibar, some amount of condensation
is still observed. What helps the most in reducing condensation, as shown in Fig.\ref{fig:Fig9},
is the mounting of a metal shield around the sample with through holes for optical beams. This is
because, a large fraction of the residual gases condense on the shield and do not
reach the sample inside. Figure \ref{fig:Fig10} shows two representative scans
of the transient reflectivity curve of the QW sample at 50 K after the cryostat
is conditioned and with the metal shield in place. It is clear that with the
slow oscillations in linear reflectivity suppressed, the transient reflectivity
signal becomes repeatable and reliable.

\section{Conclusion}
We have shown that the periodic oscillations observed in transient
reflectivity signal measured by the pump-probe technique is caused by a film
slowly condensing on its surface. In transient reflectivity measurements
based on pump-probe technique, the evolution of the reflectivity is built
up by taking repeated observations at different pump-probe delays and
therefore the full measurement can take tens of minutes to complete. Thus
even a film with a growth rate as slow as a few nm/min can cause modulations
in the measured ultrafast transient reflectivity signal. We believe that
the reflectivity oscillations observed in earlier reports could be due to
this film growth phenomenon \cite{oscillationsmanganite, Znosurfacecharge}.
Our studies on different samples at various temperatures and
the reduction in growth rate with time shows that the source of the film
formation is due to the residual gases inside the chamber.
The measures required for proper conditioning of cryostat for low-temperature
transient spectroscopy experiments are also presented. We also believe that
maintaining a vacuum level better than $1\times10^{-6}$ millibar inside the
chamber will reduce the amount of residual gases and will improve the
reliability of the measured transient reflectivity signal. Since the process of
film growth is independent of materials, the reported precautions are important
for reflection based experiments with any sample.

\section{Acknowledgment}
The authors would like to acknowledge the support from Dr. H. S. Rawat, Head,
Laser Physics Applications Section and Dr. S. C. Mehendale. The authors also
acknowledge Mrs. Asha Singh for the help during the experiments.

\bibliographystyle{unsrt}

\clearpage

\begin{figure}
\includegraphics[width=0.85\textwidth]{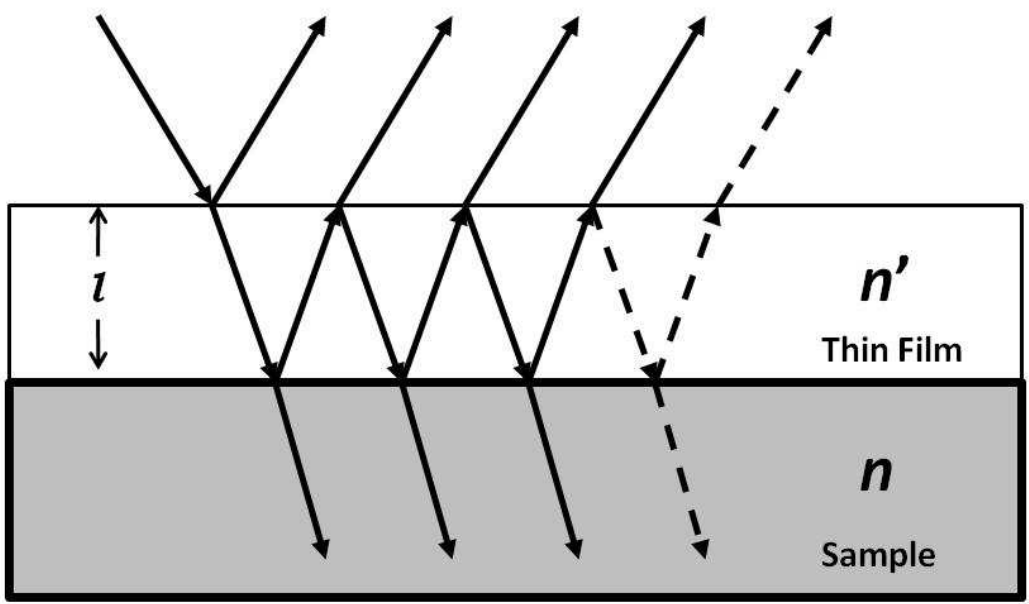}
\caption{\label{fig:Fig6} Schematic of multiple reflection from a thin film-substate
system. $l$ is the thickness of thin film. $n'$ and $n$ are the refractive index
of the thin film material and the sample respectively.}
\end{figure}

\begin{figure}
\includegraphics[width=0.85\textwidth]{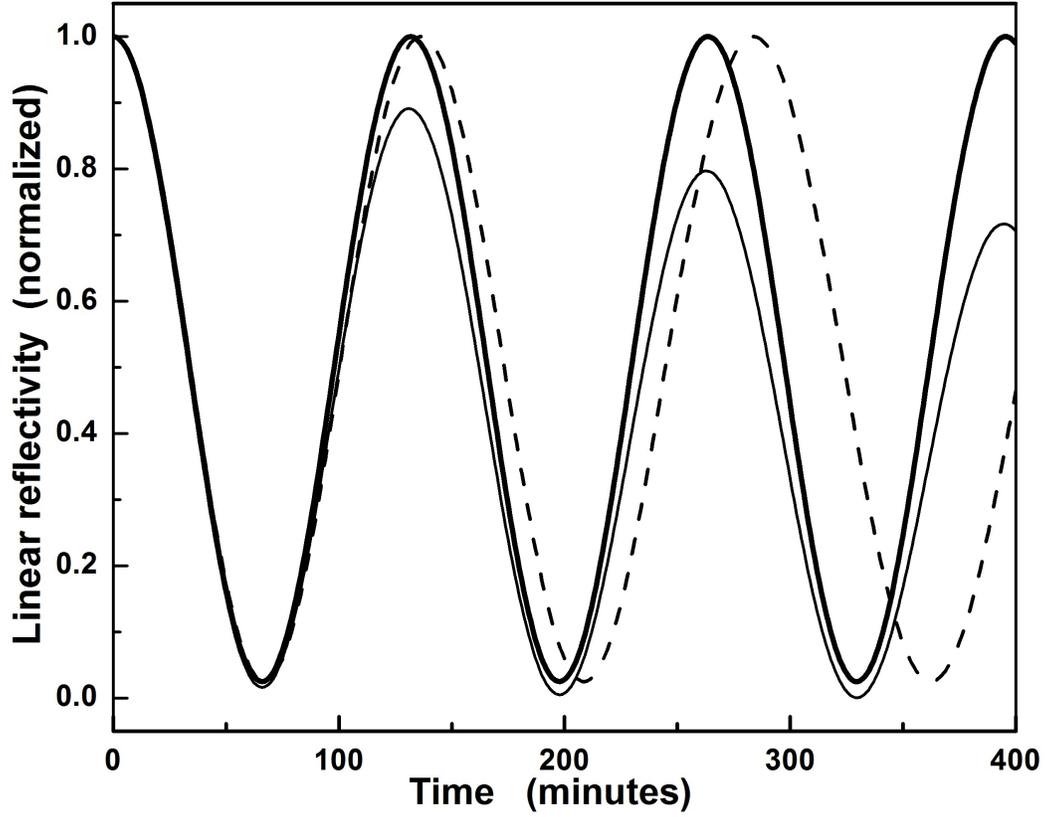}
\caption{\label{fig:Fig7} The calculated variation of linear reflectivity from
a sample and a thin film on its top surface with time using Eq.\ref{Eq:ThinFilm}:
thick line - constant $g$ and no absorption in the film ($\alpha = 0$), thin line - constant
$g$ with finite absorption in the film and dashed line - linearly decreasing
$g$ and $\alpha = 0$.}
\end{figure}

\begin{figure}
\includegraphics[width=0.85\textwidth]{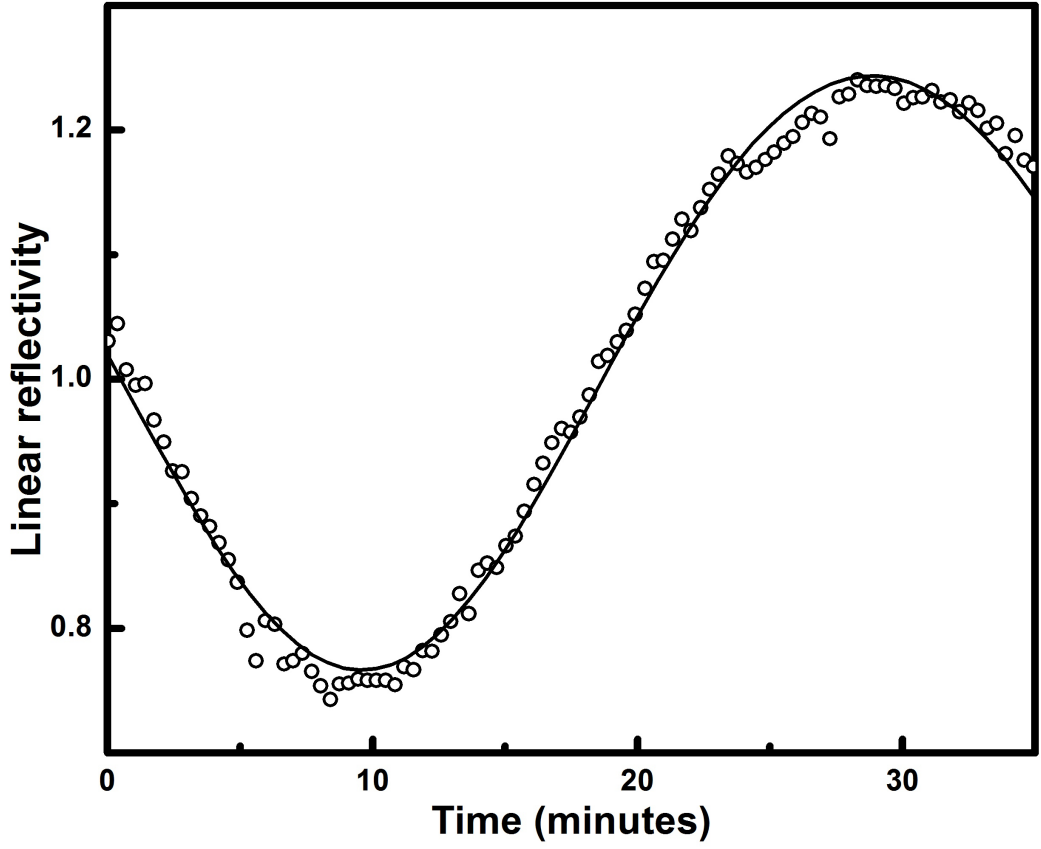}
\caption{\label{fig:Fig8}The variation of the measured linear reflectivity of bulk
GaAs sample. The calculated curve is the best fit with Eq. \ref{Eq:ThinFilm}
neglecting any absorption in the film. For ease of viewing, the reflected
signal has been normalized by dividing it with its corresponding mean value.
The refractive index of the GaAs used in fitting as well as
the best fit parameters are given in Table \ref{t1}.}
\end{figure}

\begin{figure}
\includegraphics[width=0.85\textwidth]{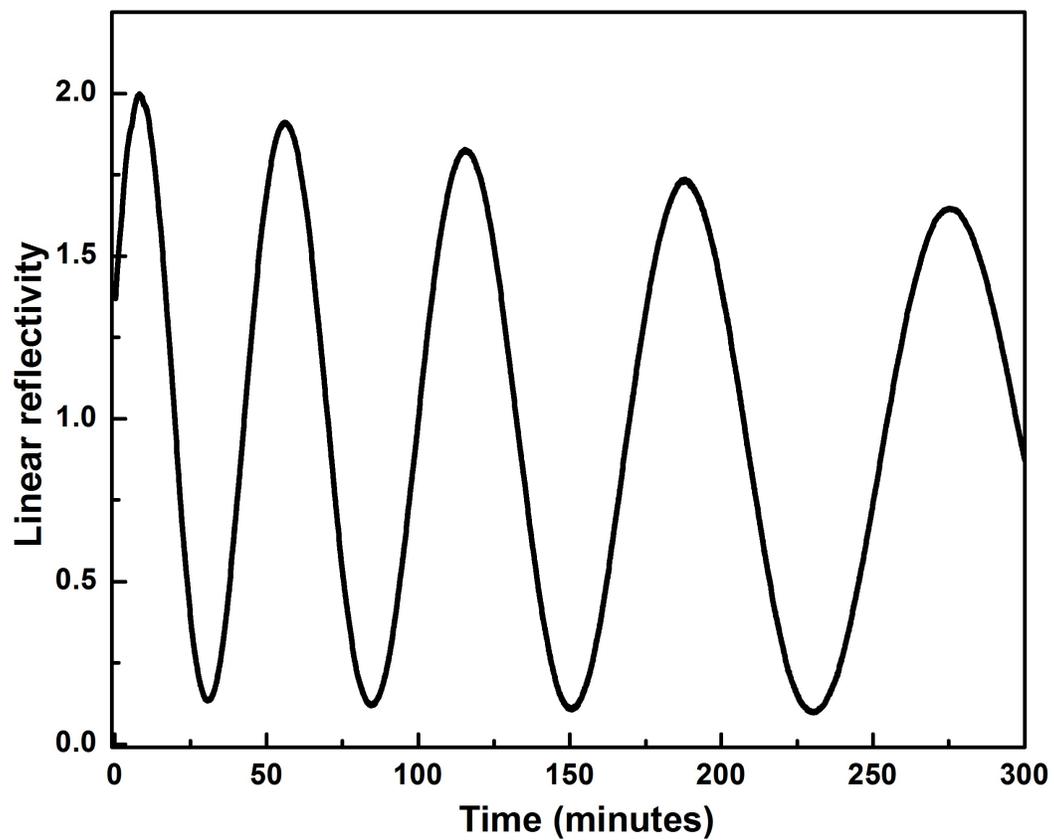}
\caption{\label{fig:Fig4} Measured time dependence of the linear reflectivity of
a glass plate in a optical cryostat at temperature 50 K in much longer time of
observations. For ease of viewing, the reflected signal has
been normalized by dividing it with its corresponding mean value.}
\end{figure}

\begin{figure}
\includegraphics[width=0.85\textwidth]{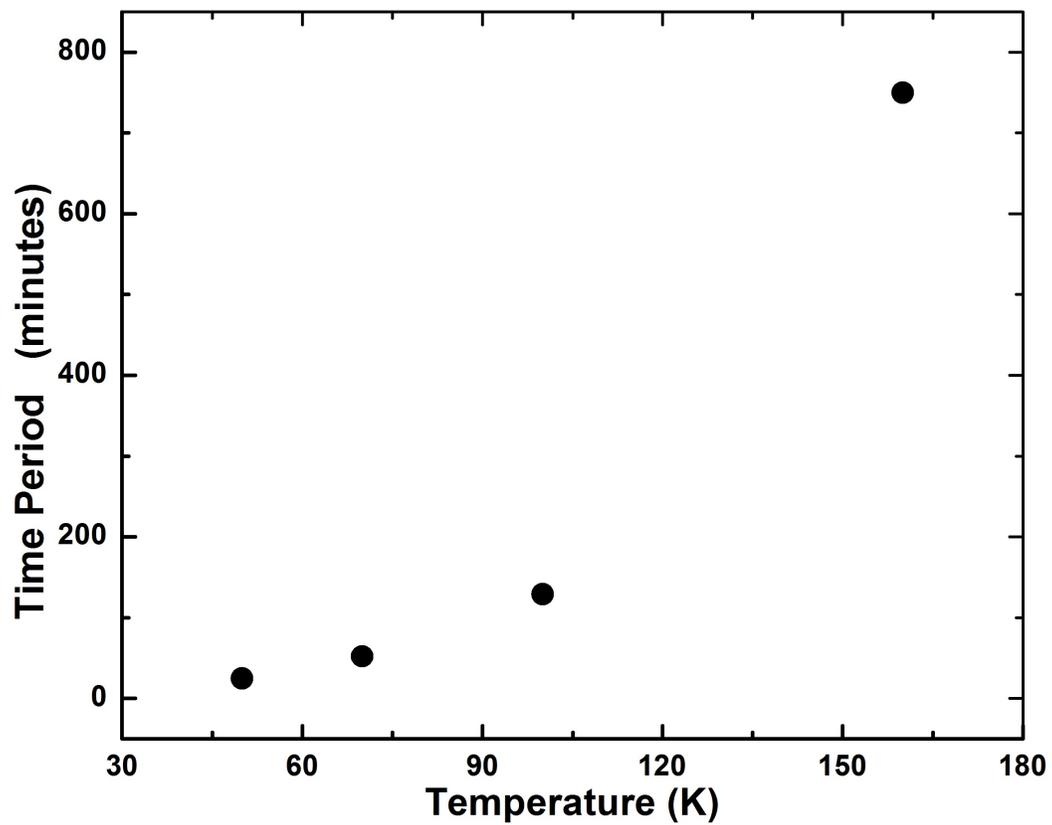}
\caption{\label{fig:Fig5} The temperature dependence of period of linear
reflectivity oscillations measured from bulk GaAs.}
\end{figure}

\begin{figure}
\includegraphics[width=0.85\textwidth]{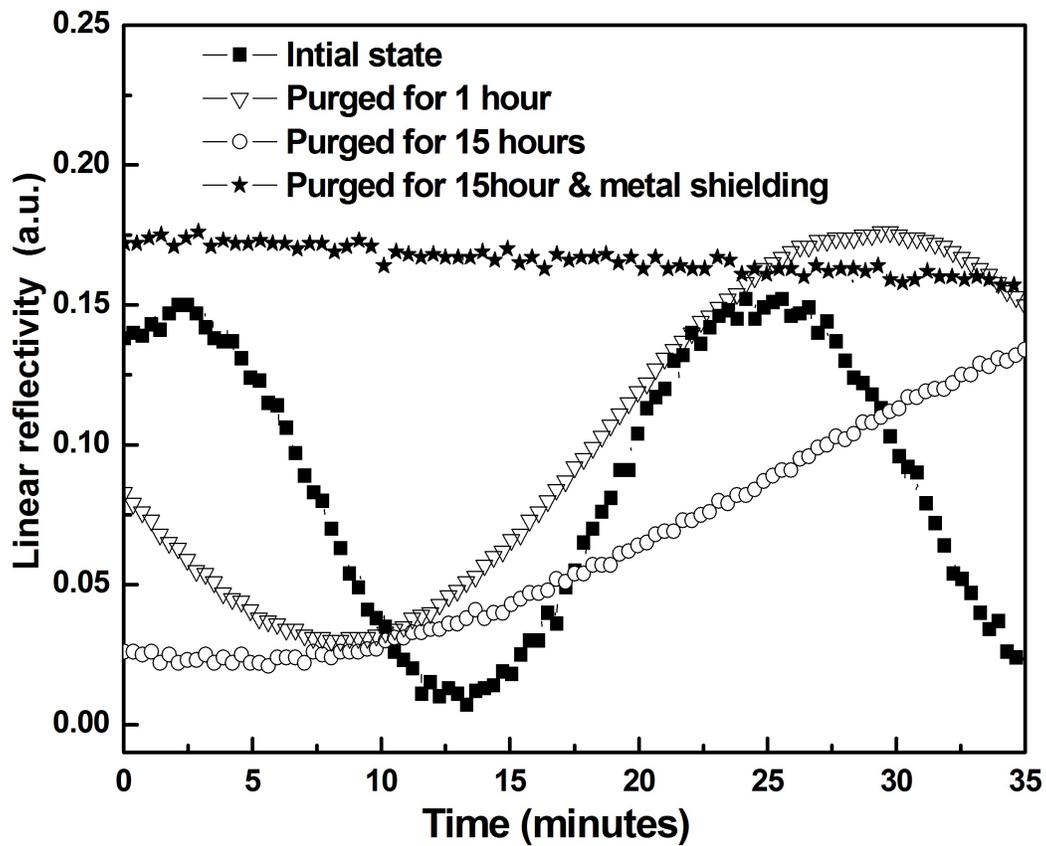}
\caption{\label{fig:Fig9} Reflectivity signal for glass at 50 K under
different nitrogen purging times as well as with an additional metal shielding
around the sample.}
\end{figure}

\begin{figure}
\includegraphics[width=0.85\textwidth]{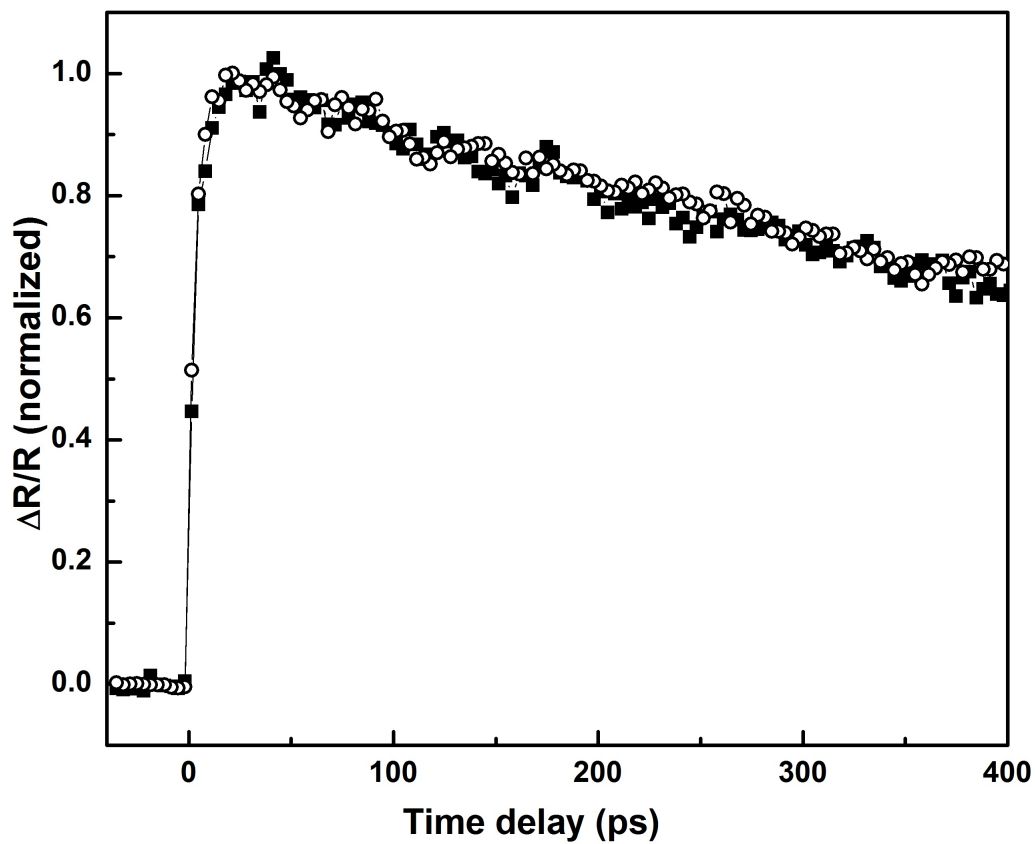}
\caption{\label{fig:Fig10} Two representative scans of transient reflectivity of
GaAsP/AlGaAs quantum well at 50 K after conditioning the cryostat and with the
metal shield a around the sample.}
\end{figure}

\clearpage

\begin{table}
\caption{The estimated values of refractive index $n'$ and growth rate of the
film condensing on different materials. The refractive index of the material used
in the calculations are also given. The wavelength is 632.8 nm.}
\label{t1}
\begin{center}
\begin{tabular}{llll}
\hline
\multicolumn{1}{c}{Material} & \multicolumn{1}{c}{$n$ }& \multicolumn{1}{c}{$n'$ }
& \multicolumn{1}{c}{$g$ (nm/minutes)} \\
\hline
\verb|Aluminium| & 0.04& 1.38&2.5 \\
\verb|Glass| & 1.54& 1.34 & 8 \\
\verb|GaAs| & 3.3& 1.15&7 \\
\hline
\end{tabular}
\end{center}
\end{table}

\end{document}